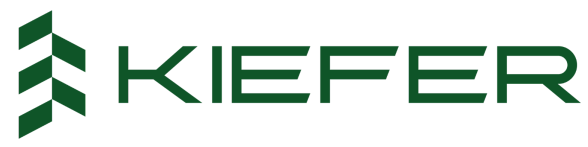
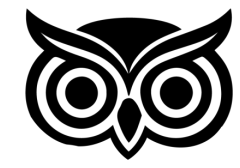

# Teaching Nemotron Greek: Mining a Corpus, Adapting Retrieval, and Grounding Generation for Modern Greek across Specialist Domains

**Ayoub Kirouane**[1] **Christos Petrocheilos**[1]

[1]Sophea AI, KIEFER SA, Athens, Greece
{a.kirouane, c.petrocheilos}@kiefer.gr
models@sophea.ai

**Abstract**

Modern Greek is missing from the supported-language list of NVIDIA's Nemotron retrieval models and from every major multilingual retrieval benchmark, yet Greek legal, energy, financial and clinical documents are exactly the long, jargon-dense text that retrieval-augmented generation is meant to serve. We adapted the Nemotron family to Greek end to end, mining the corpus, training the retrieval stack, synthesising the reader supervision and building an evaluation benchmark, then measured what each stage actually buys. Building the data surfaced findings of its own: **there is essentially no native Greek instruction data to train on**, so any pool at useful scale must be translated, and Greek document chunks are long enough that a `max_len` of 512 silently truncates **87%** of training pairs. The modelling headline is equally uncomfortable: **on all five of our domains a BM25 lexical baseline, with no learned parameters, outscores every off-the-shelf multilingual embedder we tested**, including an 8B one. Fine-tuning a 1B embedder on 65,773 Greek retrieval pairs lifts nDCG@10 from 0.362 to **0.835**, and the competence it gains generalises: on a general-domain Greek corpus that neither model was trained on, the adapted embedder beats its own unadapted base by **+0.399**. Its advantage over BM25 does not generalise in the same way. We lead in the domains we adapted on and lose on general Greek, and we report both. A cross-encoder whose off-the-shelf contribution is inconsistent across domains becomes a reliable gain once adapted. Finally, LoRA-tuning a 30B-A3B mixture-of-experts model as a grounded reader raises judged answer correctness from 29.4% to **66.9%**. We also report the four ways our own instruments lied to us along the way, including two claims of ours that a second, larger evaluation did not reproduce. We release the models and HERA, a new Greek retrieval-augmented generation benchmark, at huggingface.co/KIEFERSA.

## 1. Greek falls off the map

Dense retrieval is supposed to have superseded lexical search. That belief is load-bearing: it is why production RAG stacks ship a multilingual embedder and drop BM25 (Robertson and Zaragoza, 2009) entirely.

It does not survive contact with Greek. Greek is absent from the supported-language lists of the Nemotron retrieval models, and from BEIR (Thakur et al., 2021) and MIRACL (Zhang et al., 2023), the benchmarks that shape the field's sense of what multilingual retrieval can do. What happens to a language that is out of distribution for both the models and the yardsticks is simply not measured.

So we measured it. But first we had to build the data, and that turned out to be half the story.

## 2. What already exists

Greek is not untouched. **Meltemi-7B** (Voukoutis et al., 2024) and **Llama-Krikri-8B** (Roussis et al., 2025) adapt open models to Greek through continued pretraining and instruction tuning, and **GR-NLP-TOOLKIT** (Loukas et al., 2025) supplies Greek tokenisation, lemmatisation and tagging. That work targets general Greek generation; ours targets retrieval and grounded reading, so the artifacts do not overlap. We do not compare against those models, and that is a real gap rather than a scoping decision: a Greek-adapted reader is the control that would separate our recipe from our choice of base model.

Greek grounded reading is likewise not new. **Belebele** (Bandarkar et al., 2024) covers Greek passage-grounded multiple choice and is one of the benchmarks in Table 7, and **XQuAD** (Artetxe et al., 2020) provides Greek extractive items over a gold passage. What we could not find, and what Section 4 builds, is a Greek benchmark carrying *multi-passage* context, mined distractors, citation targets and unanswerable items. That is the narrow claim, and it is the only one we make.

Retrieval is where the absence is real. BEIR (Thakur et al., 2021) is English-only and MIRACL (Zhang et al., 2023) excludes Greek, so the multilingual embedders we test were trained and measured without it. BEIR's own conclusion, that BM25 is a robust out-of-domain baseline, is what Section 5 reproduces in a new language with the parameter axis held fixed; we claim the measurement, not the phenomenon. Multilingual E5 (Wang et al., 2024) is the obvious further baseline and we did not run it. For RAG evaluation, RAGAS (Es et al., 2024) and ARES (Saad-Falcon et al., 2024) supply metric frameworks rather than data, which is the half we had to build.

## 3. Mining the corpus

We started from 407,053 raw `(query, document, label, domain)` pairs and ended with 65,773 clean per-query records across five domains: energy, legal, finance, medical and a general foundation slice. The reduction is not aggressive filtering for its own sake; most of it is regrouping flat pairs by query. But one cleanup step is worth naming, because mined corpora reliably need it. We call it **positive-wins**: a document mined as a hard negative for a query, which is in fact that query's own positive, gets dropped. Negative mining produces this constantly, and training on it teaches the model to push apart two things it should pull together.

The queries are synthetic, and *how* they were synthesised matters more than the fact that they were. Left to itself, an instruct model writes short, keyword-like queries, which look nothing like what real users type: long, formal, fact-seeking questions after a specific decision number, date, capacity or amount. A retriever trained on keyword queries scores well on synthetic evaluations and then stumbles in production, so the generation was constrained rather than free. Two generators wrote roughly three questions per chunk: **Nemotron-3-Ultra-550B** for the general foundation slice and **GLM-5.2** for the four style-matched domain slices. Three controls applied throughout: each call is few-shot conditioned on real production user queries so the register transfers; a strict grounding rule requires the specifics a query asks for to appear in its own chunk, audited at 97–99% with few-shot exemplar leakage near zero; and difficulty is varied deliberately rather than left to chance, which Section 4 pushes furthest with an explicit L1–L5 ladder.

Labelling rests on one bet: **the chunk a query was written from is that query's positive**. No human marks relevance anywhere in the corpus. The bet is cheap to state and easy to check, and the audit above is what checks it: if 97–99% of queries ask for specifics that appear in their own chunk, then the chunk answers the query by construction.

**Negatives are mined, and the miner matters more than the sampler.** A reranker only learns from contrast, so negatives are the dominant quality lever. An off-the-shelf Qwen3-Embedding-8B embeds queries and passages asymmetrically, the query side carrying a retrieval instruction and the document side plain, and per query we take a cosine-kNN window over that query's *own-domain* pool, so an energy query gets energy distractors rather than a trivially wrong legal one. Two guards define the window: skip the top few ranks, because those are too likely to be genuinely relevant and would become false negatives, and drop anything above a similarity cap, because those are near-duplicates of the positive. Around six negatives per query survive. Mining the same queries with BM25 instead of a dense embedder produced measurably worse negatives, so the lexical signal that makes BM25 a strong *retriever* on Greek (Section 5) does not make it a good *negative miner*. The result is separable but hard: on a sampled audit, query-to-positive similarity averaged 0.687 against 0.474 to the mined negatives, a margin of 0.213, with no query weakly grounded to its own positive.

**Style-matching is worth a lot, and synthetic evaluation still flatters.** Both effects were measured directly on production queries. Real users write long, multi-part questions averaging about 176 characters, roughly a quarter of them exact identifier lookups, against about 68 characters for naive synthetic queries; few-shot style-matching the synthetic set to real ones *doubled* the reranking lift over BM25, from +0.040 to +0.081 nDCG@10. The sobering half is an in-house observation we report as such: on real user queries the margin over BM25 was roughly **+0.04**, against +0.080 on our synthetic held-out set, so under half the lift survived contact with production traffic. That comparison is unpublished, and we attach no query count or labelling protocol to it; it is an internal observation, not a measured result. Synthesised queries echo their source passage's vocabulary and are therefore systematically easier than what users type. Every retrieval number in this report is measured on synthetic held-out queries. They are sound for ranking systems against each other, which is what we use them for, but they are an upper bound on production behaviour.

We then tagged language by Greek-script ratio (Greek characters over Greek plus Latin, computed *after* stripping code blocks so that technical passages are not misread as English), decontaminated train against the union of every evaluation query, deduplicated by normalised exact match, and split by query with seed 42, verifying zero overlap.

A `max_len` of 512, the most common default in retrieval tooling, truncates roughly **87%** of our training pairs. This was the single most consequential thing the data told us. Greek document chunks are long: positives have a median of 1,196 characters (≈1,034 tokens), p95 of 1,288 and a maximum of 4,407, while queries are short at a median of 113 characters. Both models are therefore trained at `max_len` 4,096. Anyone reproducing this at 512 would train on truncated positives and conclude the method does not work.

The corpus is deliberately bilingual, 87.1% Greek, with English retained at 12.9% rather than filtered out, because code-switched Greek/English is pervasive in Greek technical and legal writing. Both models are trained bilingually in a single run, with no per-language head or adapter.

### 3.1. Synthesising a reader that cites

No gold answers exist for a retrieval corpus, so reader supervision had to be manufactured. We built 40,000 examples from the retrieval *train* split only, keeping it disjoint from every retrieval test query. The teacher is **Sophea-Titan-1**, our Greek-adapted 27B model (a LoRA on Qwen3.6-27B, released at huggingface.co/KIEFERSA/Sophea-Titan-1). It answers each query using *only* the gold passage; if it cannot, the row becomes an abstention example. The context is then assembled as the gold passage plus $k \in [2, 5]$ shuffled hard negatives, numbered, with the gold landing at a random position $p$, and the target is `answer [p]`, an answer carrying an explicit citation.

The point of that construction is that robustness is *designed in rather than hoped for*. The distractor count varies, gold position is shuffled specifically to counteract the lost-in-the-middle effect (Liu et al., 2024), and abstention is an explicit training target rather than an emergent behaviour. Roughly 20% of examples are abstentions built from negatives alone.

**The audit caught our own judge.** Auditing the synthesised corpus put grounded answer faithfulness at 95.2%, with 3 wrong citations in 30,903 and zero duplicate queries. Abstention label noise, however, measured an alarming 20.4%. Before accepting that, we ran a control in which the gold passage was *provably* removed, rows that must be abstentions, and the same judge scored those at 23.5%. The control came in *worse* than the real data, which means 20.4% was mostly the judge's own false-positive floor rather than corpus noise. Real abstention noise is bounded near 5%.

We also name three gaps the audit exposed rather than smoothing over: all 30,903 grounded answers cite exactly *one* chunk, so there is no multi-document citation supervision; abstention targets are drawn from only two canned strings; and 8.7% of answers are ultra-terse.

### 3.2. Native Greek data barely exists

**There is essentially no commercially licensed, human-written Greek instruction data at scale.** The volume that exists is machine-translated or templated: the Greek slice of the Aya collection (Singh et al., 2024; Üstün et al., 2024) runs to millions of rows that way, while its human-annotated Greek portion is on the order of hundreds, and the open assistant corpora we surveyed carry no Greek at all. Anything at useful scale therefore has to be translated.

Our Greek instruction pool is 296,034 conversations, of which the filtered, deduplicated split used downstream is 206,909 training rows (the figure that reappears in Section 4 and Section 9), and assembling it produced the scarcity finding that constrains everything downstream. At this scale a Greek instruction pool is *necessarily* translation-heavy, which carries a translationese risk that supervised fine-tuning will shape directly into model behaviour. This is not a solved problem and we do not claim to have solved it. It is the honest reason we cannot cleanly separate a base-model ceiling from a data ceiling.

Everything in the pool passes the same filter: non-empty turns, Greek-ratio thresholds computed after code-stripping, length bounds, degeneracy detection, exact-signature deduplication, and MinHash near-duplicate removal at 0.8 both within the pool and against it. That filter retains 92.1% of the augmentation set.

## 4. A Greek benchmark, because none existed

Greek grounded reading has been evaluated only through translated single-passage resources (Section 2), and no Greek benchmark we could find puts a reader in front of many passages at once, so the evaluation had to be built before anything could be measured. We built **HERA** (Hellenic Retrieval-Augmented), a 4,946-item Greek Wikipedia long-retrieval benchmark with citations (huggingface.co/datasets/KIEFERSA/HERA), summarised in Table 1, and it is the yardstick behind every reader number in this report. Appendix C defines the metric set it is scored with.

Table 1: The Greek Wikipedia long-retrieval benchmark. Source is Greek Wikipedia under CC-BY-SA-4.0, released with attribution and revision identifiers.

| property | value |
|---|---|
| items | 4,946 |
| answerable | 3,712 |
| unanswerable | 1,234 (25%) |
| multi-hop | 1,145 (23%) |
| difficulty | L1 factoid to L5 synthesis |
| context | 8 to 40 passages, 21.2 mean |
| construction | Qwen3.5-122B generates, GLM-5.2 verifies |
| human pass | spot-check of random samples |
| decontamination | 13-gram check, **zero** overlap |

Two design choices matter. First, the difficulty ladder and the deliberate 25% share of *unanswerable* items mean the benchmark measures abstention as a first-class skill, not as an afterthought. Most RAG benchmarks score only whether an answerable question was answered well, which cannot detect a model that confidently invents an answer when the context does not contain one.

Second, HERA is built in two stages. **Qwen3.5-122B** generates the question, answer and citation; **GLM-5.2** then independently verifies that the item is answerable only from its source passage, correct, and well formed. We picked Qwen3.5-122B for the generator role because its Greek is markedly fluent for a general multilingual model, which is the property that matters most here: a benchmark whose questions and reference answers read as translationese would test the wrong thing, and Section 3.2 has already established how hard native Greek text is to come by. We report that as the qualitative basis for the choice, not as a measured result. Only cross-validated items survive, and a random sample was additionally spot-checked by hand. Exhaustive human review of 4,946 long-context items was out of scope, which is the reason the second model stage exists at all. Hard distractors are mined separately, by hybrid BM25 plus dense retrieval fused with RRF and then reranked, so the context noise is the highest-scoring non-gold passage rather than a random one.

**A confound we have to declare.** Our reader judge is a served Qwen3.5-122B-A10B-FP8, the same family as the model that authored HERA's items. The judge is therefore grading text from its own family throughout, and family preference is a documented failure mode of LLM-as-judge (Zheng et al., 2023). Two things push against it: no item survives without independent verification by GLM-5.2, a different family, and the spot-check sampled the corpus by hand. Neither removes the effect, and we did not measure it.

What it does *not* touch is worth stating, because the exposure is uneven. Only *judged answer correctness* is scored against a model-authored reference answer. Faithfulness is scored against the retrieved context, the citation metrics against the gold passage index, and abstention against items whose gold passage was removed by construction. Those three are anchored to the benchmark's structure rather than to any model's phrasing, and they carry the largest gains. Answer correctness is the most exposed number here.

On human verification we claim only what we can support. Labels are two strong models' consensus plus a hand spot-check of random samples. We did not log how many items were checked, so we report no coverage figure and do not present it as systematic validation. A sized, logged human audit remains the most valuable single addition to this evaluation.

**One thing we got wrong.** The generator over-provisioned candidates three-to-one against a 5,000-item target, expecting strict verification to reject many. **We failed to log the realised acceptance rate.** We can bound it only at $\geq 1/3$ and cannot say how strict the gate actually was. It is the weakest link in our reader evaluation, and we would rather flag it than infer a number.

Decontamination, at least, was verified rather than assumed. A 13-gram overlap check between the benchmark (77,411 distinct normalised 13-grams) and the full 206,909-row instruction pool returns **zero** overlapping rows. Normalisation strips accents first, so Greek orthographic variation cannot defeat the check. The benchmark is also out of corpus for the reader, whose training data comes from energy, legal, finance and medical documents, so it measures transfer rather than in-domain fit.

Every retrieval number below is macro nDCG@10 over 5,830 held-out queries, and every comparison carries a paired bootstrap interval. For systems $A$ and $B$ scored per query on a query set $Q$, we draw $Q_b$ from $Q$ with replacement and recompute the mean difference,

$$\hat{\Delta}_b = \frac{1}{|Q_b|} \sum_{q \in Q_b} (s_A(q) - s_B(q)),$$

for $b = 1, ..., B$ with $B = 10,000$ and seed 42; the 95% interval is the 2.5th and 97.5th percentiles of $\left\{\hat{\Delta}_b\right\}$. The pairing matters: $A$ and $B$ are differenced on the *same* query before averaging, which removes per-query difficulty from the variance and is why intervals as tight as $[+0.072, +0.089]$ are attainable at this sample size.

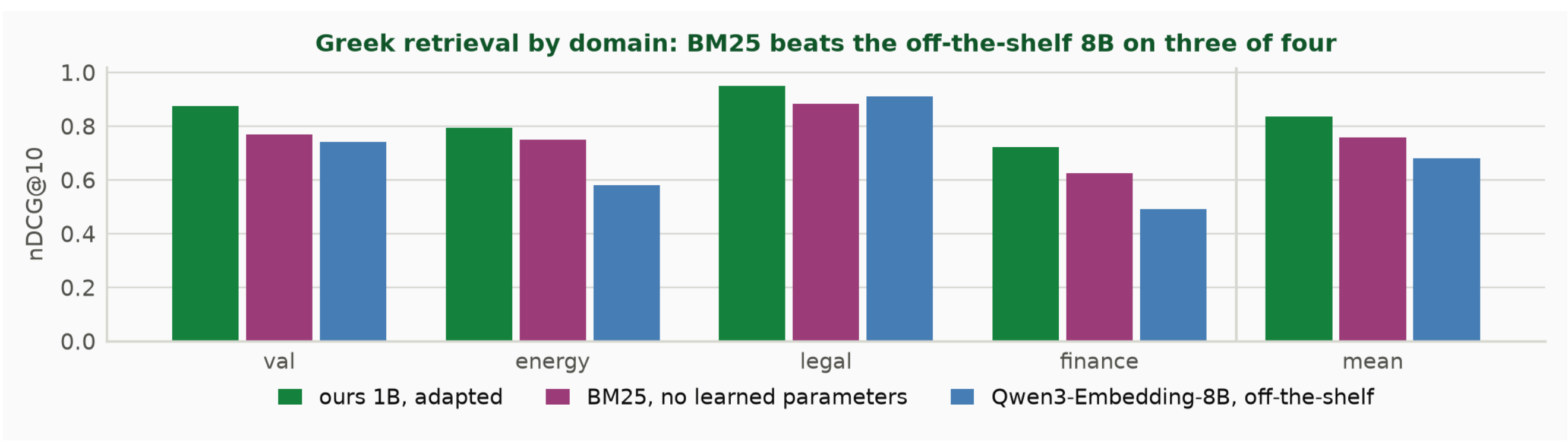

Figure 1: Greek retrieval, four domains plus the macro mean, 5,830 held-out queries. A parameter-free lexical baseline beats a state-of-the-art 8B multilingual embedder on three of the four sets and loses one. Adapting a 1B model beats both.

Table 2: The numbers behind Figure 1 and Figure 2. Full held-out IR over 5,830 queries, corpus = all positives. Δ is our adapted 1B against the unadapted base, with a paired bootstrap interval over queries; no interval approaches zero. The three right-hand columns are the off-the-shelf Qwen3-Embedding family, unadapted. Medical is not a row here: it has its own corpus and so cannot share this table's protocol, and is reported separately in Section 6.

| set | queries | base | ours 1B | Δ [95% CI] | BM25 | Qwen 8B | Qwen 4B | Qwen 0.6B |
|---|---|---|---|---|---|---|---|---|
| val | 2,000 | 0.3823 | **0.8755** | +0.4933 [.475,.514] | 0.7694 | 0.7402 | 0.7383 | 0.6579 |
| energy | 1,122 | 0.3454 | **0.7937** | +0.4483 [.423,.475] | 0.7505 | 0.5804 | 0.5786 | 0.5089 |
| legal | 1,769 | 0.5257 | **0.9497** | +0.4241 [.401,.444] | 0.8840 | 0.9096 | 0.9187 | 0.8629 |
| finance | 939 | 0.1935 | **0.7218** | +0.5283 [.503,.555] | 0.6246 | 0.4902 | 0.5015 | 0.4265 |
| macro over sets | 5,830 | 0.3617 | **0.8352** | +0.4735 | 0.7571 | 0.6801 | 0.6843 | 0.6140 |
| macro over queries | 5,830 | 0.3883 | **0.8575** | +0.4692 | 0.7772 | 0.7206 | 0.7242 | 0.6542 |

## 5. The uncomfortable baseline

Figure 1 is the result we did not expect.

**BM25 scores 0.757. Qwen3-Embedding-8B scores 0.680. The unadapted Nemotron embedder scores 0.362.**

A lexical baseline with no learned parameters, no GPU and no training beats a state-of-the-art multilingual dense embedder on Greek. It wins on energy (+0.170), finance (+0.134) and general validation (+0.029), and loses only legal (−0.026). We do not read that loss as legal Greek being closer to a multilingual model's distribution. Legal is the easiest set for *every* system we ran, and by each one's own standard: BM25 peaks there at 0.8840, the unadapted Nemotron at 0.5257, our adapted 1B at 0.9497. A domain that is uniformly easy is not evidence about any one family. Greek legal queries turn on article numbers, dates and statute references, which anchor lexically whatever is doing the ranking.

The practical reading is blunt: *any Greek RAG system that replaced BM25 with an off-the-shelf multilingual embedder made its retrieval worse on these domains.* The scope of that sentence is deliberate. Every number in this section is in-domain, and the out-of-domain counterpart below is weaker: there BM25 stays ahead of the 0.6B and 4B rungs, but is no longer separable from the 8B.

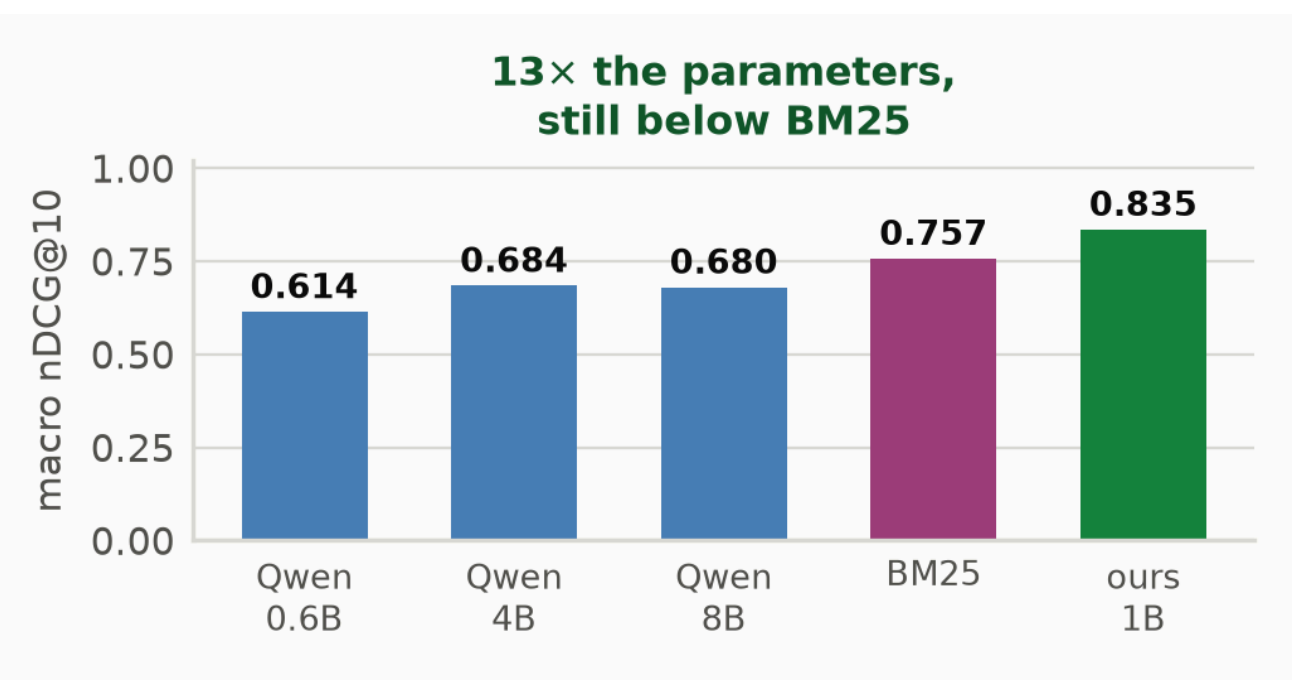

Figure 2: The off-the-shelf embedder family across a 13× parameter range, in-domain. Every rung sits below BM25; 4B and 8B differ by −0.004. Out-of-domain the same three rungs separate cleanly (Table 4).

Nor does scale rescue it (Figure 2). Within the same off-the-shelf family (Zhang et al., 2025), same training and same protocol, differing only in size, 8B and 4B are indistinguishable (0.6801 vs. 0.6843, a gap of −0.004), and dropping all the way to 0.6B costs only 0.070 (0.6140). The entire family sits below BM25 across a 13× parameter range. On this corpus, doubling the parame-ter count buys nothing, because the missing ingredient is not capacity. It is exposure to the language.

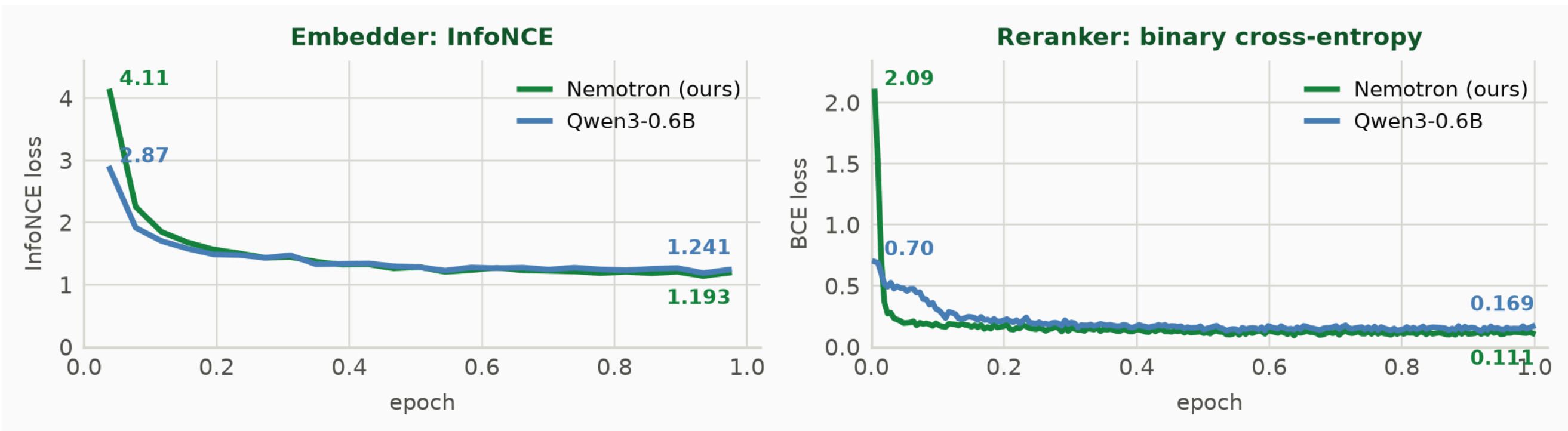

Figure 3: Training loss over one epoch, both stages, against the same recipe run on Qwen3-0.6B. Within a panel the two runs share objective, corpus, schedule and logging interval. The Nemotron models start markedly higher and finish lower: Greek was already supported by the Qwen backbone before training, and out of distribution for Nemotron. Absolute values are not comparable across panels (InfoNCE against BCE), nor strictly across backbones, which differ in tokenizer and parameterisation.

**Where that claim stops.** We later ran the same three rungs out of domain, on the HERA retrieval track (the protocol is Table 4; we give these three here because that table is matched to the size of *our* models). Off domain the family is not flat at all. It is a clean monotone ladder, $0.5268 \rightarrow 0.6182 \rightarrow 0.6504$, every step significant, $+0.1237$ end to end across the same 13× range, and the 8B rung finally catches BM25: $+0.0062$, 95% CI $[-0.0057, +0.0177]$, not separable. So the flatness above is a property of *this corpus*, not of Greek retrieval, and so is the clean sweep over the whole family. Capacity does buy something on general Greek text; it bought nothing on specialist Greek text, where exposure was the binding constraint. We state the narrow version because we can only support the narrow version: one family, two query distributions.

## 6. A 1B model, adapted, wins

We fully fine-tuned Nemotron-3-Embed-1B (NVIDIA, 2025a) contrastively with InfoNCE (Oord et al., 2018). For a batch of $B$ anchors, query $q_i$ is scored by cosine similarity $s$ against a candidate set $\mathcal{C}_i$ holding its own positive $p_i^+$, every other in-batch positive (Karpukhin et al., 2020), and all $7 \times B$ mined hard negatives, seven per query:

$$\mathcal{L}_{\text{emb}} = -\frac{1}{B}\sum_{i=1}^{B} \log \frac{e^{s(q_i, p_i^+)/\tau}}{\sum_{d \in \mathcal{C}_i} e^{s(q_i, d)/\tau}}$$

The denominator is the whole point: it is where the $7 \times B$ hard negatives live, so the number of negatives, and therefore the quality of the gradient, is set by the *logical* batch rather than by what fits in memory. GradCache (Gao et al., 2021) decouples the two, letting us hold a logical batch of 256 at 4,096 tokens on a single GPU. Both stages start well above the Qwen control and finish below it (Figure 3): Greek is inside that backbone's distribution, outside Nemotron's. Familiarity is not competence; it still lost to BM25.

Macro nDCG@10 goes from 0.362 to **0.835** (Table 2, which reports every domain with its interval; Appendix A defines the metrics). Every domain gains at least +0.42, and the gain is largest exactly where the base was weakest: finance +0.53, starting from 0.19. A base score under 0.21 does not mean degraded. It means non-functional on Greek financial text.

Against BM25 the margin is $+0.080$, 95% CI $[+0.072, +0.089]$, paired bootstrap over queries. The interval is nowhere near zero. A 1B model that has seen the language beats both a parameter-free baseline and an 8B model that has not.

**Medical.** Medical is evaluated on its own set: 650 queries over 24,812 passages, drawn from the same public source as the rest of the medical corpus. It has its own corpus, which is why it is reported here rather than as a row of Table 2. Table 3 is the result, and it is the medical number we report anywhere in this paper.

Table 3: Medical, 650 queries over 24,812 passages. This is the medical evaluation we report; it needs its own corpus, which is why it is not a row of Table 2.

| system | nDCG@10 | R@10 |
|---|---|---|
| Nemotron-3-Embed-1B, base | 0.1586 | 0.2292 |
| **ours 1B, adapted** | **0.6867** | **0.8815** |
| BM25 (parameter-free) | 0.6602 | 0.8554 |
| Qwen3-Embedding-8B | 0.5936 | 0.7892 |
| Qwen3-Embedding-4B | 0.6119 | 0.8046 |
| Qwen3-Embedding-0.6B | 0.4698 | 0.6354 |

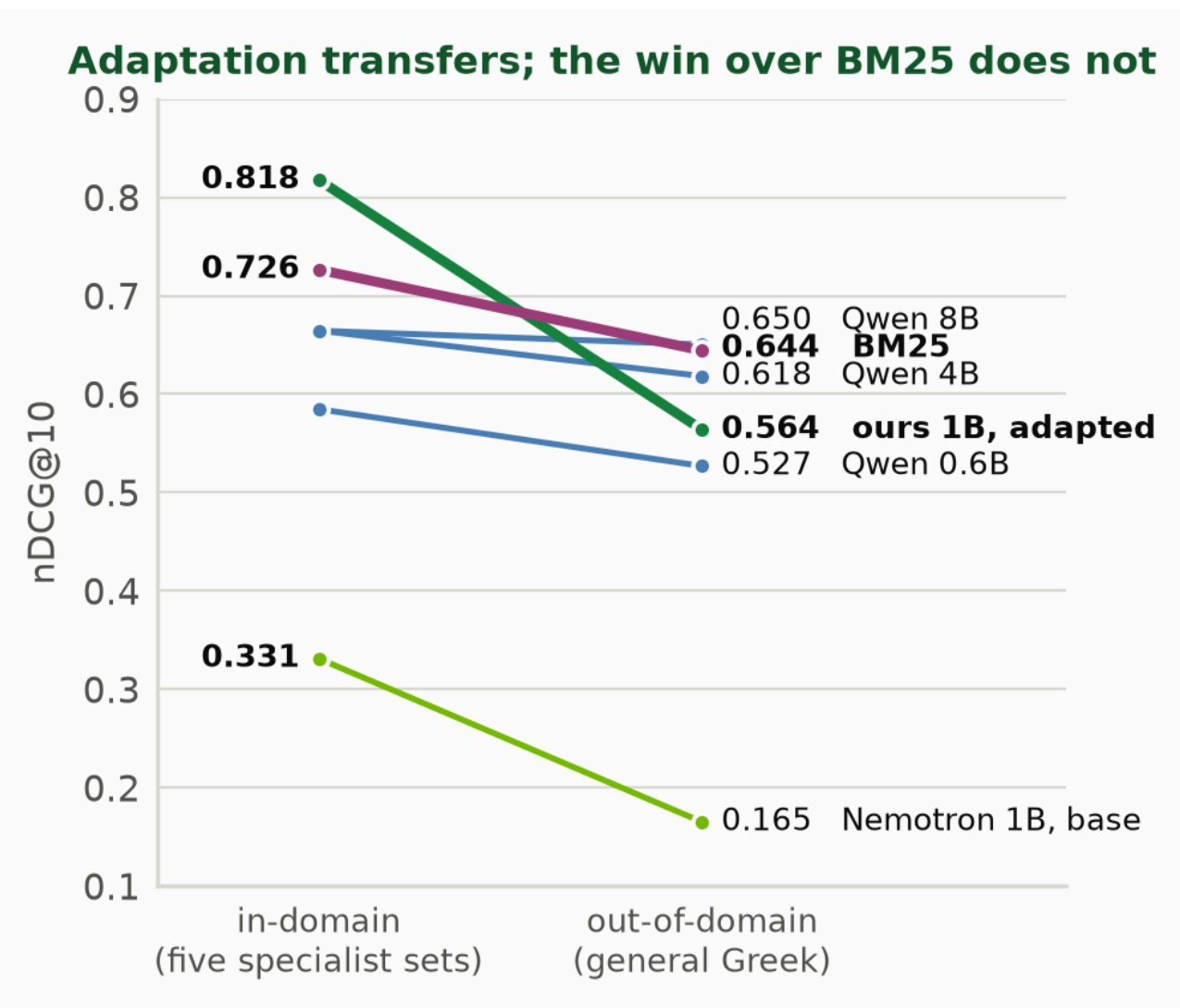

Figure 4: The same systems in domain and out of it. Levels are *not* comparable across the two evaluations, which use different corpora; the ordering is the point. Our adapted 1B leads in domain and falls behind BM25 and two Qwen rungs outside it, while staying far above the base it was built from. The three Qwen rungs share one colour and are told apart by their labels.

Two things hold there and one does not. **Adaptation holds, overwhelmingly**: +0.528 over the unadapted base, 95% CI $[+0.496, +0.561]$. **The lexical baseline holds too**: BM25 at 0.6602 still beats every off-the-shelf embedder, including the 8B at 0.5936 and the 4B at 0.6119 (+0.048, CI $[+0.020, +0.076]$), so Section 5′s central finding survives here as well. What does not hold is our margin over BM25. At +0.027, 95% CI $[-0.0005, +0.054]$, **the interval touches zero**: on this set we cannot demonstrate an advantage over lexical search at this sample size.

**Out of domain: what the adaptation actually transferred.** Everything above is measured on the domains we adapted on. To see what survives outside them we ran the same models on the HERA retrieval track (Figure 4, Table 4): 4,946 queries against 300,000 Greek Wikipedia passages, general-domain text that neither the base models nor ours were trained on. Two things separate cleanly, and they point in opposite directions.

**The adaptation transfers as language competence.** Out of corpus, our 1B beats the unadapted Nemotron it was built from by +0.399, 95% CI $[+0.387, +0.410]$, and our 0.6B beats its own base by +0.049, CI $[+0.041, +0.056]$. Teaching a model Greek made it better at Greek retrieval generally, not only at ours. That is the cleanest evidence in this paper for *exposure, not capacity*: the 1B base sat at 0.165, barely functional, and adaptation moved it to within 0.007 of a 0.6B model whose backbone already spoke Greek.

Table 4: Out of domain: the HERA retrieval track, 4,946 queries over 300,000 Greek Wikipedia passages, a corpus none of these models trained on. *base* is the off-the-shelf checkpoint. Compared at matched size, with BM25 as the parameter-free reference. Every adapted-vs-unadapted gap here is significant; both adapted models sit *below* BM25 alone, and the fusion of BM25 with our 1B sits above it.

| system | params | nDCG@10 | R@10 |
|---|---|---|---|
| BM25 (parameter-free) | — | 0.6442 | 0.7326 |
| **BM25 + ours 1B, RRF** | 1B | **0.6715** | **0.7823** |
| Qwen3-Emb-0.6B, base | 0.6B | 0.5268 | 0.6189 |
| Qwen3-Emb-0.6B, **ours** | 0.6B | 0.5753 | 0.6728 |
| Nemotron-3-Emb-1B, base | 1B | 0.1651 | 0.2105 |
| Nemotron-3-Emb-1B, **ours** | 1B | 0.5637 | 0.6611 |

**The win over lexical search does not transfer.** BM25 scores 0.6442 here and beats both our adapted models, by +0.081 over the 1B and +0.069 over the 0.6B, and off-the-shelf Qwen3-Embedding-4B beats our 1B by +0.055. The ordering we report in-domain reverses. We take this as a measured boundary rather than a caveat: the phrase *across specialist domains* in our title is doing real work, and a reader deploying this embedder on general Greek should expect to lose to a lexical baseline. One property of both evaluations limits how far this settles the question. HERA's queries, like ours, are LLM-generated from their gold passages, which shares vocabulary between query and gold and flatters lexical matching on *both* sides. That common bias cannot explain the reversal between them, but it does mean BM25′s absolute standing is likely generous in each.

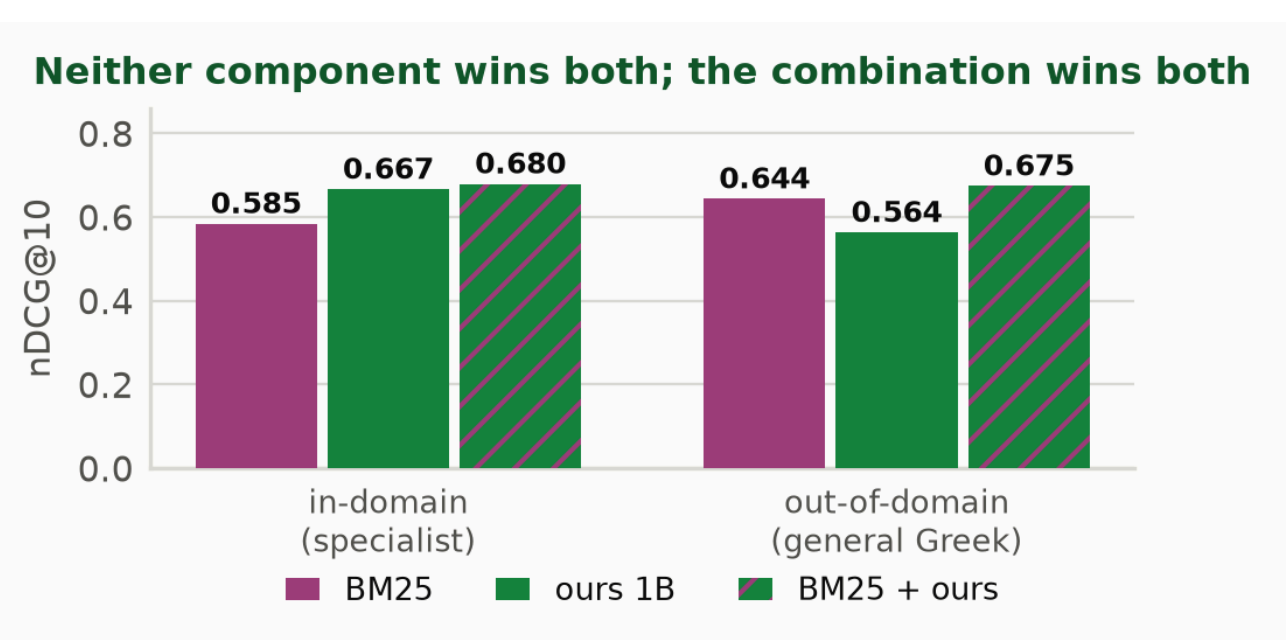

Figure 5: Dense and lexical retrieval fused by reciprocal rank fusion. The hatched bar is not a third system: it is the two beside it combined. Neither component wins both regimes and the combination wins both. The fusion weight is chosen on a held-out fold and scored on the other, so no bar is tuned on the queries it is scored on.

**The right move is to add BM25, not to replace it.** A baseline that beats you is a component, not only a rival, so we fused the two with reciprocal rank fusion (Cormack et al., 2009), $\mathrm{RRF}(d) = \sum_i w_i/(k + r_i(d))$ at $k = 60$ over each system's top 100. **With equal weights and nothing tuned, the fusion scores 0.6715 out of domain against BM25's 0.6442** (Figure 5), a gain of +0.027, 95% CI [+0.019, +0.035], and it lifts Recall@10 from 0.733 to 0.782. Letting a weight be chosen on a held-out fold and scoring only the other fold reaches 0.6749, which is not separable from the untuned version, so the result does not depend on tuning at all.

The same holds in domain, where the dense model is the stronger component rather than the weaker one. On a 2,500-query in-domain sample our 1B alone scores 0.6670 and BM25 0.5846; the fused system reaches 0.6798, +0.013 over the dense model alone, CI [+0.007, +0.019], with the weight again chosen on a held-out fold. The weight moves in the direction you would expect, favouring the dense side in domain and the lexical side outside it, which is a small piece of evidence that the gain is complementarity rather than a fitting artifact. The same holds for our other embedder: fusing the fine-tuned Qwen3-0.6B with BM25 gains +0.034 out of domain, CI [+0.028, +0.041], and +0.016 in domain, CI [+0.010, +0.022], over whichever of its two components is stronger there. Both of our embedders behave the same way, so the effect is a property of combining dense with lexical retrieval on Greek rather than of one checkpoint.

**So the honest conclusion is not that a lexical baseline beats our embedder.** It is that neither system dominates, that they fail differently, and that the combination beats the better of the two in *both* regimes: +0.013 where our model leads and +0.027 where BM25 does. For a Greek RAG system the practical recommendation follows directly, and it is not the one we expected when we started: run both.

**Our other fine-tuned embedder, finally compared.** We also ship a fine-tuned Qwen3-0.6B embedder. In-domain the two have still never been comparable, because that run used a harder corpus with positives *plus all* negatives as distractors. Table 4 is the first time they have been scored head to head under a single protocol, and out of domain the 0.6B is ahead: 0.5753 against 0.5637, a difference of +0.012, 95% CI [+0.004, +0.020]. Separable, but small enough that the honest summary is that 40% fewer parameters cost nothing here. Which model to deploy is a question about your corpus, not about their size.

**It also shrinks.** The embedding retains Matryoshka structure (Kusupati et al., 2022) through fine-tuning

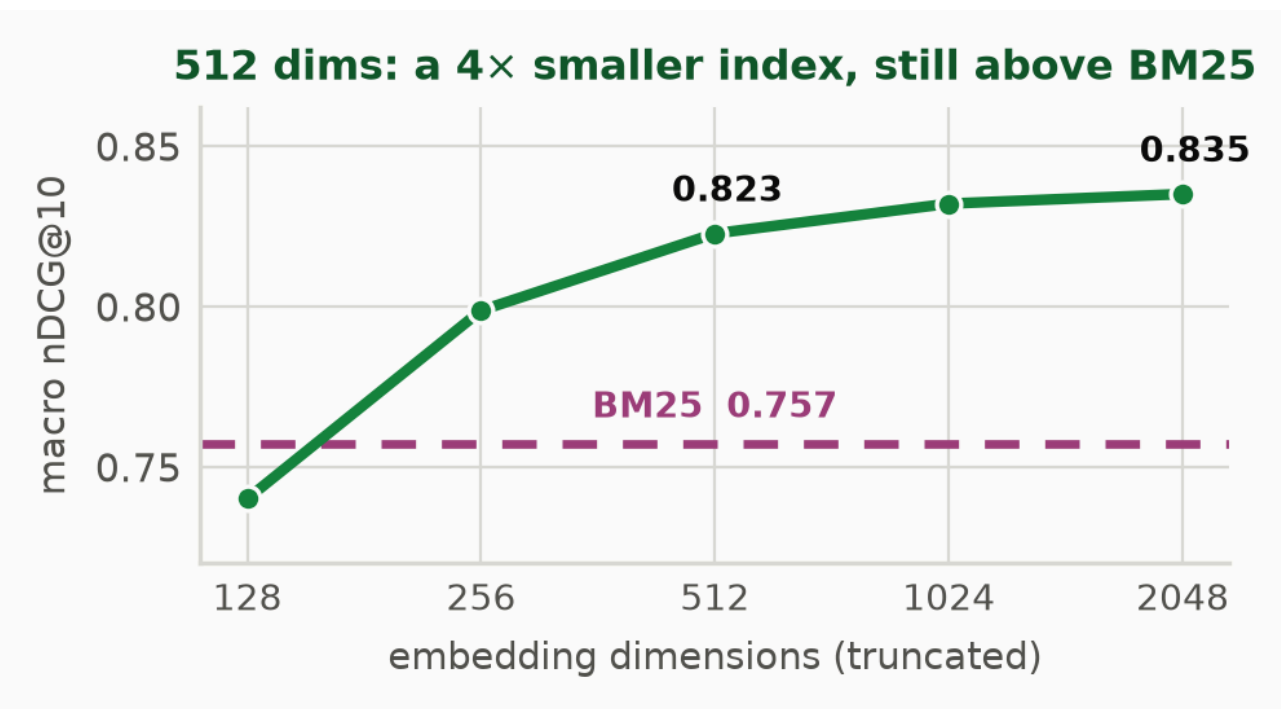

Figure 6: Matryoshka truncation. Quality is flat down to 512 dimensions and only collapses below BM25 at 128.

(Figure 6). Truncated to 512 dimensions the index is 4 × smaller and still scores 0.823, which is 98.5% of full quality and still +0.066 over BM25.

## 7. The second stage, measured against its floor

The reranker is a cross-encoder (Nogueira and Cho, 2019) trained pointwise rather than contrastively. Each canonical row expands into labelled pairs, $y = 1$ for a query's positive and $y = 0$ for each of up to eight negatives, giving 394,579 pairs $\mathcal{P}$ at roughly 11% positive. A single relevance logit $f(q, d)$ is fit with binary cross-entropy, writing $\hat{y} = \sigma(f(q, d))$:

$$\mathcal{L}_{\mathrm{rr}} = -\frac{1}{|\mathcal{P}|} \sum_{(q,d,y)\in\mathcal{P}} [y \log \hat{y} + (1 - y) \log(1 - \hat{y})]$$

Pointwise scoring is what lets the second stage see a query and document together, which the first stage never does.

Table 5 gives the per-domain result, and Appendix B records what it was measured on: the fixed first stage here is our fine-tuned Qwen3-0.6B embedder over 750 pooled queries, so this floor is not a row of Table 2. Second stages are usually assumed to help. We measured that assumption against the right control: the no-rerank floor, i.e. the first stage's own ranking.

On this evaluation **the off-the-shelf reranker's contribution is statistically indistinguishable from not reranking at all** (−0.006, $p = 0.52$), while adapted, the same architecture is worth +0.047 ($p < 0.001$). Per domain the off-the-shelf arm significantly degrades one set (validation, −0.050, $p = 0.018$) and significantly helps another (legal, +0.039, $p = 0.003$), which averages to nothing.

**We then measured it again, and one of those conclusions did not hold.** A second-stage result that rests on 750 queries is worth re-running, so we repeated the comparison on **2,580** queries over the same five domains, with the same fixed first stage and the same top-50 candidate depth (Figure 8, Table 6). The two runs differ in more than one respect: query sample, per-set corpora, and the medical set of Section 6. We therefore do not attribute the difference between them to any single cause, and we take as established only what holds in both.

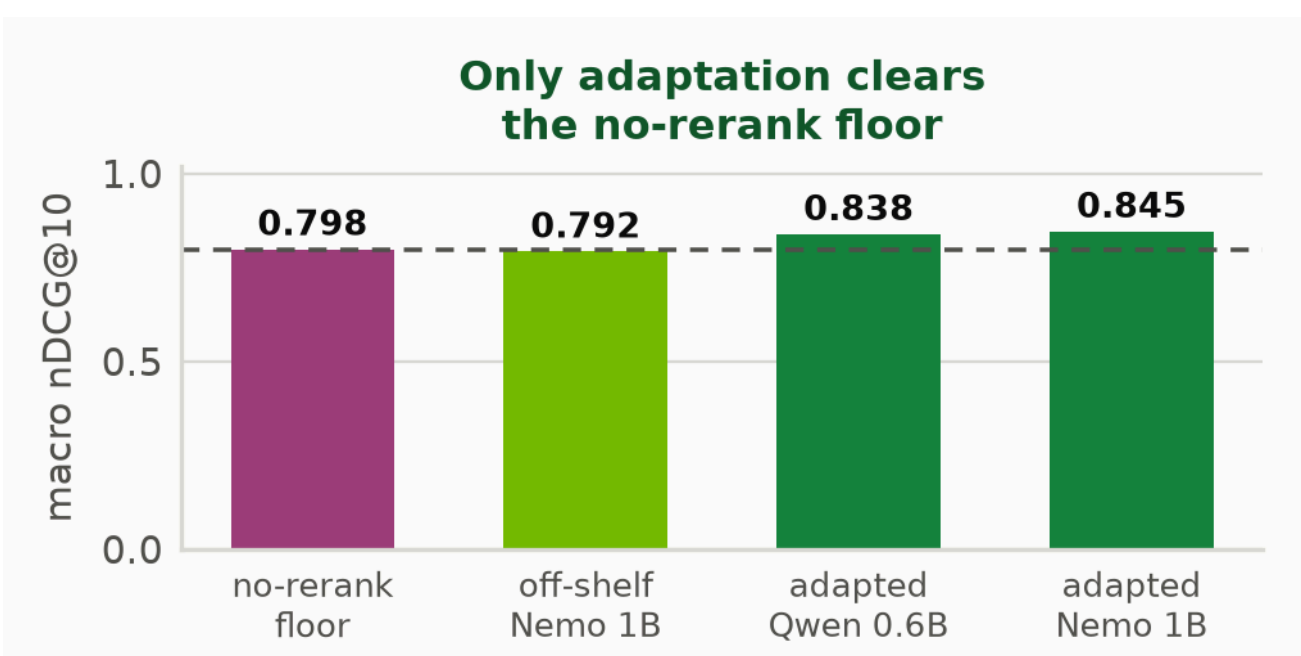

Figure 7: Reranking against the no-rerank floor (dashed), first stage held fixed at top-50, 750 pooled queries. Off-the-shelf: $-0.006$, CI $[-0.026, +0.013]$, $p = 0.52$. Adapted Nemotron: $+0.047$, CI $[+0.028, +0.066]$. The two adapted arms are not separable: $+0.007$, CI $[-0.006, +0.019]$, $p = 0.29$.

Table 5: The numbers behind Figure 7, per domain. 150 queries per domain, 750 pooled. *floor* is the first stage's own ranking, *base* the off-the-shelf Nemotron cross-encoder, *ours* our adapted Nemotron-1B, *Qwen* our adapted Qwen3-0.6B. Bold marks the best in each row. Every row is scored against its own corpus, and the medical row uses the set of Section 6.

| set | floor | base | ours | Qwen |
|---|---|---|---|---|
| val | 0.8438 | 0.7936 | 0.8740 | **0.8800** |
| energy | 0.7746 | 0.7412 | 0.7800 | **0.7841** |
| legal | 0.9264 | 0.9655 | **0.9692** | 0.9614 |
| finance | 0.6737 | 0.6975 | 0.7548 | **0.7557** |
| medical | 0.7729 | 0.7621 | **0.8485** | 0.8104 |
| mean | 0.7983 | 0.7920 | **0.8453** | 0.8383 |

**What holds in both.** Adaptation is what makes a second stage worth its cost. Our adapted Nemotron beats the off-the-shelf cross-encoder it was built from by $+0.029$, 95% CI $[+0.022, +0.037]$ on the larger evaluation, consistent with the $+0.053$ gap between the same two arms on the smaller one, CI $[+0.035, +0.072]$.

**What does not hold.** On the larger evaluation the off-the-shelf reranker *does* clear the floor, by $+0.032$, CI $[+0.023, +0.041]$. We are no longer willing to say an unadapted cross-encoder contributes nothing. What we can say is that its contribution is *inconsistent*: it is significant on two of five sets, indistinguishable from zero on two more, and negative in point estimate on energy. Averaged over domains it is worth about half of what adaptation buys, and on the smaller sample it was worth nothing at all. A component whose benefit depends this strongly on which slice you measure is not one to add on faith.

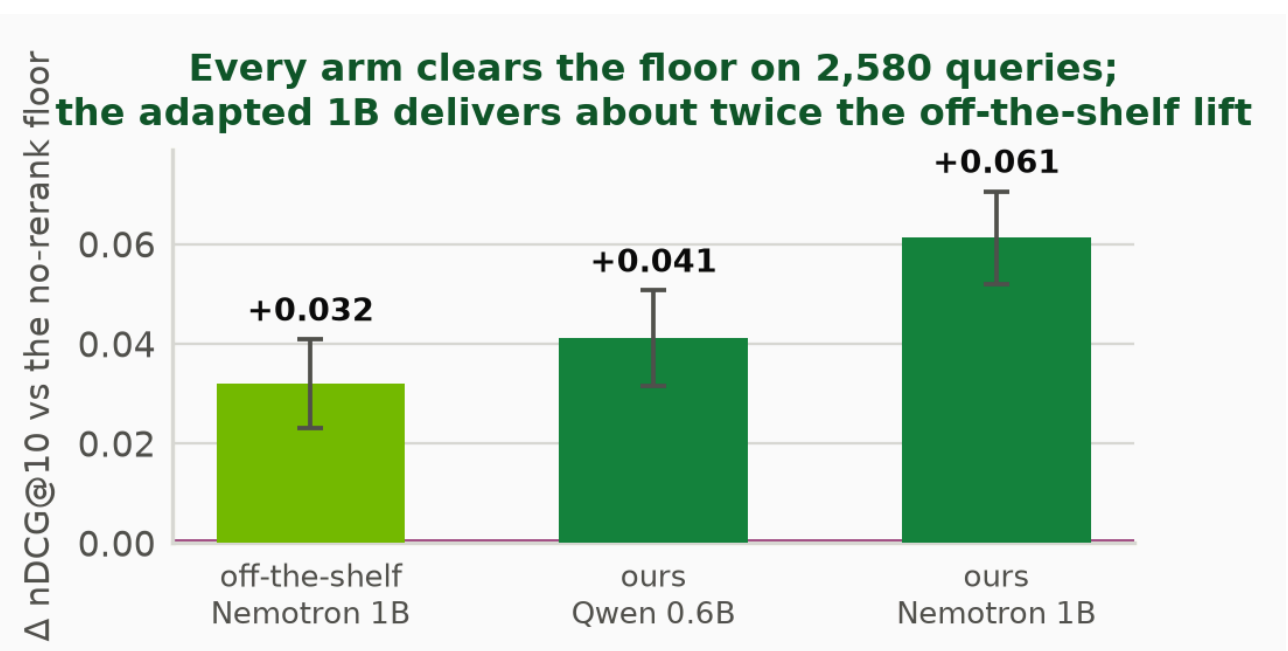

Figure 8: The second reranker evaluation, plotted as the gain over the no-rerank floor with 95% paired bootstrap intervals. Absolute nDCG@10 spans 0.82 to 0.89 here, so the difference that the section argues about is the $\Delta$, not the level. Compare Figure 7, where the off-the-shelf arm did not clear the floor at all.

The methodological point is unchanged, and is the reason we could see any of this. No evaluation that omits the no-rerank floor can distinguish these cases. Comparing reranker A to reranker B tells you which is better; only the floor tells you whether either belongs in the pipeline. It is also what told us our first answer was too strong.

**A 0.6B cross-encoder, adapted, nearly matches a 1B one.** We ran the same recipe on Qwen3-0.6B (Zhang et al., 2025), a different family at 60% of the parameters. On the 750-query evaluation the two were not separable: 0.8383 against our adapted Nemotron's 0.8453, a paired difference of $+0.007$, 95% CI $[-0.006, +0.019]$, $p = 0.29$. On 2,580 queries they do separate, and the larger model is ahead by $+0.020$, CI $[+0.014, +0.027]$. We report the separation because we found it, but the effect is small enough that the practical reading barely changes: 40% fewer parameters costs about two nDCG points, and the 0.6B still clears both the floor and the off-the-shelf 1B (which it beats by +0.009, CI $[+0.001, +0.018]$).

Table 6: The same comparison on 2,580 queries: five domains, first stage fixed, top-50 candidate depth. *floor* is the first stage's own ranking. $\Delta$ is against the floor, paired bootstrap over queries. Every arm clears the floor here, and the ordering is unambiguous.

| system | nDCG@10 | Δ | 95% CI |
|---|---|---|---|
| no-rerank floor | 0.8248 | — | |
| Nemotron-1B, off the shelf | 0.8567 | +0.032 | [+0.023, +0.041] |
| **ours**, Qwen-0.6B | 0.8659 | +0.041 | [+0.032, +0.051] |
| **ours**, Nemotron-1B | **0.8861** | +0.061 | [+0.052, +0.071] |

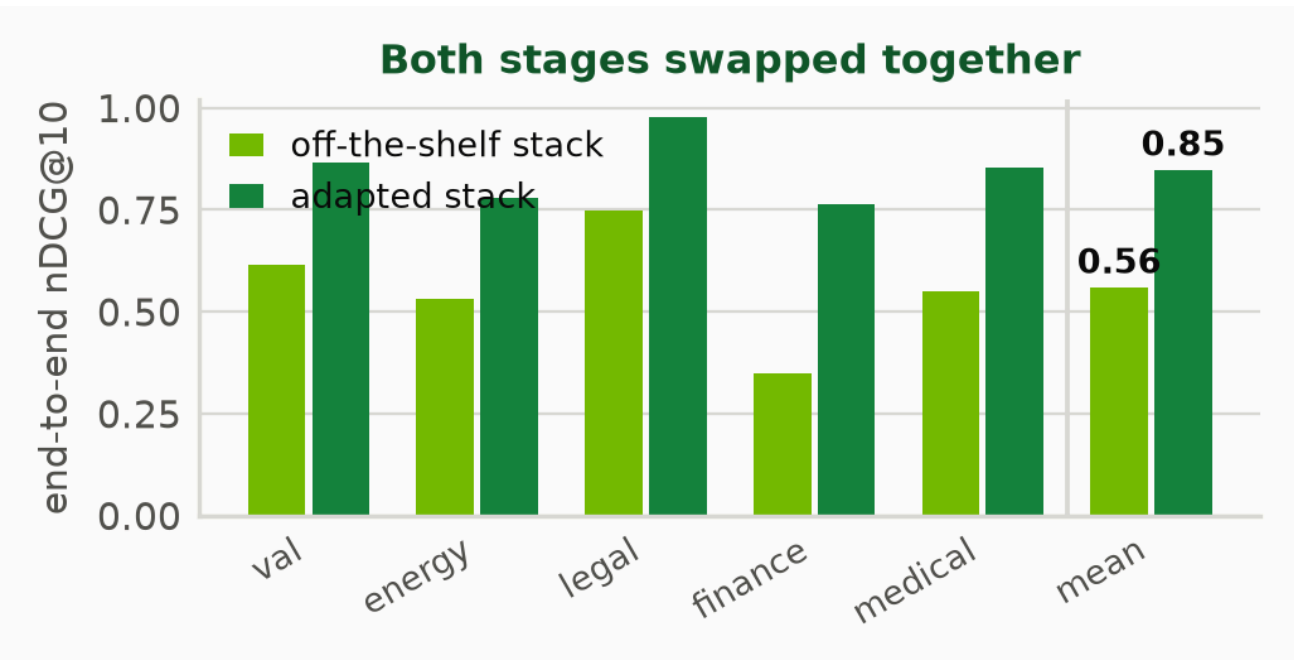

Figure 9: End-to-end two-stage retrieval, both stages swapped together.

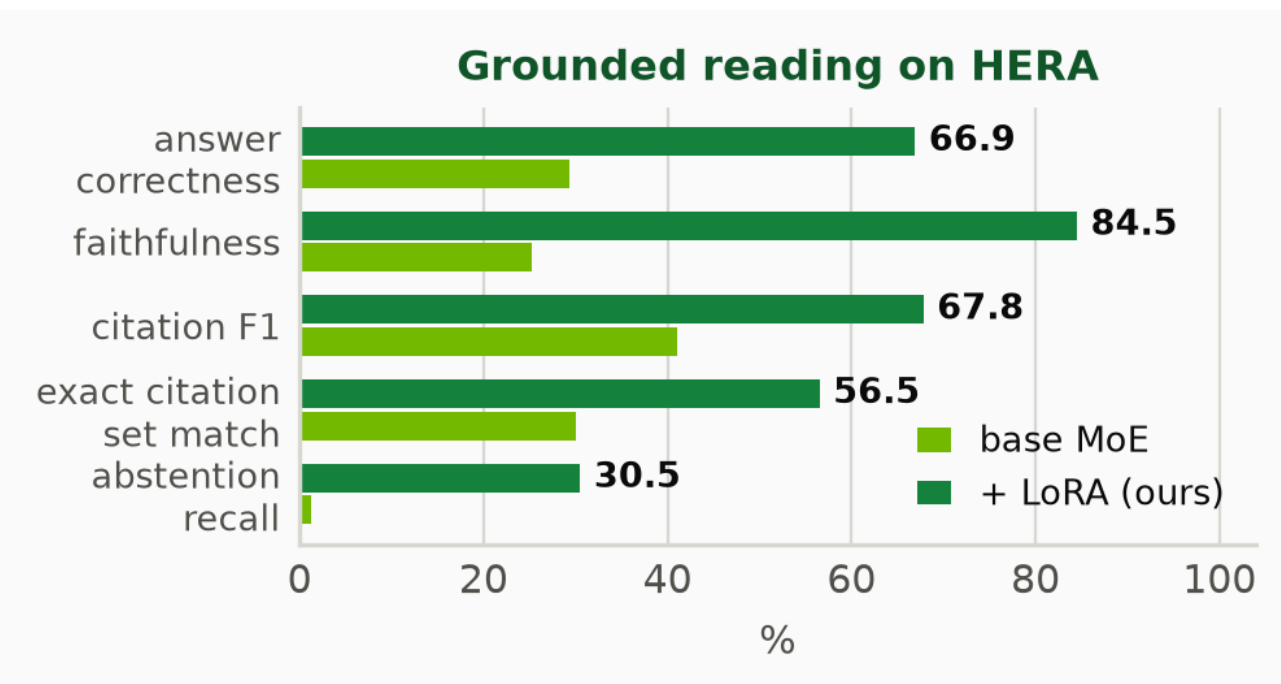

Figure 10: Grounded reading on the benchmark of Table 1. Base and adapted models receive an *identical* prompt instructing both citation and abstention, so these rows measure compliance, not whether the model was asked.

The ordering in Table 6 is the summary of this section. Both adapted rerankers sit above the off-the-shelf one, which sits above the floor. Parameter count separates the two adapted arms by a little; adaptation separates them from the unadapted one by three times as much. What the second stage was mostly missing was never architecture or scale. It was exposure to Greek.

## 8. Chained: what the reader actually receives

Swapping both stages together (Figure 9) lifts end-to-end nDCG@10 from 0.559 to **0.848**, a 52% relative gain, concentrated where the off-the-shelf stack was worst: finance +0.41, medical +0.30.

The number that decides system behaviour, though, is Recall@10, the fraction of queries whose answer is present in the context the reader actually receives. It rises from **0.624 to 0.955**. The off-the-shelf stack loses the answer entirely for about 38% of queries; ours loses it for under 5%.

This bounds *grounded* answering, not answering. A reader cannot ground an answer in a passage it never received, so writing Hit@$k$ for the fraction of queries with at least one gold passage retrieved,

$$\mathbb{E}[\text{grounded correct}] \leq \text{Hit@}k.$$

The off-the-shelf first stage therefore caps grounded end-to-end accuracy at 0.624 no matter how good the reader is, while ours raises that ceiling to 0.955. Two caveats keep this from governing Section 9. A model can be correct *ungroundedly*, reciting a fact it knew rather than one it read, and our base reader is correct more often than it is faithful (29.4% against 25.2%), which is that gap made visible. And Section 9 scores reading on gold-provided context, so those numbers are measured at Hit@$k = 1$ by construction and are not the product of this ceiling with a reader. We did not measure end to end on retrieved context, which is the number that would compose the two.

## 9. A MoE reader that cites and abstains

We LoRA-tuned (Hu et al., 2022) Nemotron-3-Nano-30B-A3B (NVIDIA, 2025b), a mixture-of-experts model (Shazeer et al., 2017; Fedus et al., 2022) with 3B active parameters, and evaluated on the benchmark of Section 4. The 40k cite-and-abstain examples of Section 3.1 are only 16.2% of a 246,909-row blend: training on them alone teaches the format at the cost of everything else, so the remainder is our Greek instruction pool, itself carrying a 7.2% English replay slice intended to limit catastrophic forgetting. Table 7 reports what that moved on benchmarks we never targeted, including where it moved the wrong way: Greek gains on six of eight while English MMLU falls 0.098 and English arc-challenge 0.094, and the mean over all thirteen is −0.004. We ran no zero-replay arm, so nothing here attributes the Greek half to the replay slice.

**The adapter.** Rank 16, $\alpha$ 32, dropout 0.05, on attention, MLP and the Mamba `in_proj`: 439M trainable parameters, 1.37% of the model. One epoch at 8,192 tokens, assistant-only loss, cosine LR $10^{-4}$, gradient clipping 0.3, effective batch 32 on four B200s, 32.7 hours.

Judged answer correctness rises 29.4% → **66.9%** (Figure 10; judge models and decoding settings in Appendix D, prompts in Appendix E); Wilson 95% intervals over the judged items are $[27.9, 30.9]$ and $[65.3, 68.4]$. But the more revealing move is faithfulness: 25.2% → **84.5%** ($[23.8, 26.7]$ and $[83.2, 85.7]$), a 3.4× gain against correctness's 2.3×. The base model does not merely answer wrongly, it answers *ungroundedly*, producing content the retrieved context does not support roughly three times in four. Since constraining generation is the entire purpose of a retrieval stage, an unadapted reader substantially wastes it.

Table 7: The *full* thirteen-benchmark sweep on capabilities neither half of the blend targets. Base and adapted are scored identically, accepting Greek and Latin answer letters on both arms. Greek improves on six of eight, English regresses on three of five, and over all thirteen the mean change is −0.004. The replay slice did not prevent forgetting; with no zero-replay arm we cannot say what it prevented.

| *Greek* benchmark | $n$ | base | adapted | Δ | *English* benchmark | $n$ | base | adapted | Δ |
|---|---|---|---|---|---|---|---|---|---|
| arc-challenge | 1,168 | 0.5445 | 0.5308 | −0.0137 | arc-challenge | 1,172 | 0.8823 | 0.7884 | −0.0939 |
| arc-easy | 1,500 | 0.6320 | 0.6587 | +0.0267 | arc-easy | 1,500 | 0.9640 | 0.8947 | −0.0693 |
| belebele | 900 | 0.6744 | 0.7089 | +0.0345 | hellaswag | 1,500 | 0.4920 | 0.5693 | +0.0773 |
| greekmmlu | 1,500 | 0.5993 | 0.5440 | −0.0553 | mmlu | 1,500 | 0.6953 | 0.5973 | −0.0980 |
| hellaswag | 1,500 | 0.3480 | 0.3973 | +0.0493 | winogrande | 1,267 | 0.6788 | 0.7001 | +0.0213 |
| medical MCQA | 432 | 0.2060 | 0.2176 | +0.0116 | | | | | |
| truthfulqa | 817 | 0.3305 | 0.3427 | +0.0122 | | | | | |
| winogrande | 1,267 | 0.5114 | 0.5604 | +0.0490 | | | | | |
| mean Δ | | | | +0.0143 | mean Δ | | | | −0.0325 |

**Correctness is conditional, and that flatters us.** Both arms are scored only on grounded items they attempted, and they attempt different numbers: the base declines 16 of the 3,712 answerable items, the adapted reader 128. Counting every refusal on an answerable item as wrong instead, correctness reads 28.7% → **64.3%** over the full 3,712, so the conditional figure credits the adapted arm with about 2.6 points it earns by refusing rather than by answering.

**Position bias disappears.** The base model degrades monotonically as the gold passage moves later in the context (31.9% → 29.4% → 26.1%), the effect reported by Liu et al. (2024). The adapted model shows no such ordering: early and middle land within 0.8 points of each other (65.8% and 65.0%). We claim the monotone penalty is *removed*, not inverted. These are single-run point estimates, and a late-vs-middle gap is not an established reversal.

**The advantage narrows as context grows.** Figure 11 tracks this. The adapted reader falls from 76.5% correct at 8 documents to 59.3% at 40, a 17-point decay, while the base model stays flat at 26–32% across the whole range. That flatness is not robustness, it is a floor effect: the base is not degrading with context because it was never using it. Faithfulness holds up better than correctness over the same range (86.1% to 79.4%), so what the long-context setting costs is mostly the ability to find the answer, not the discipline to stay grounded.

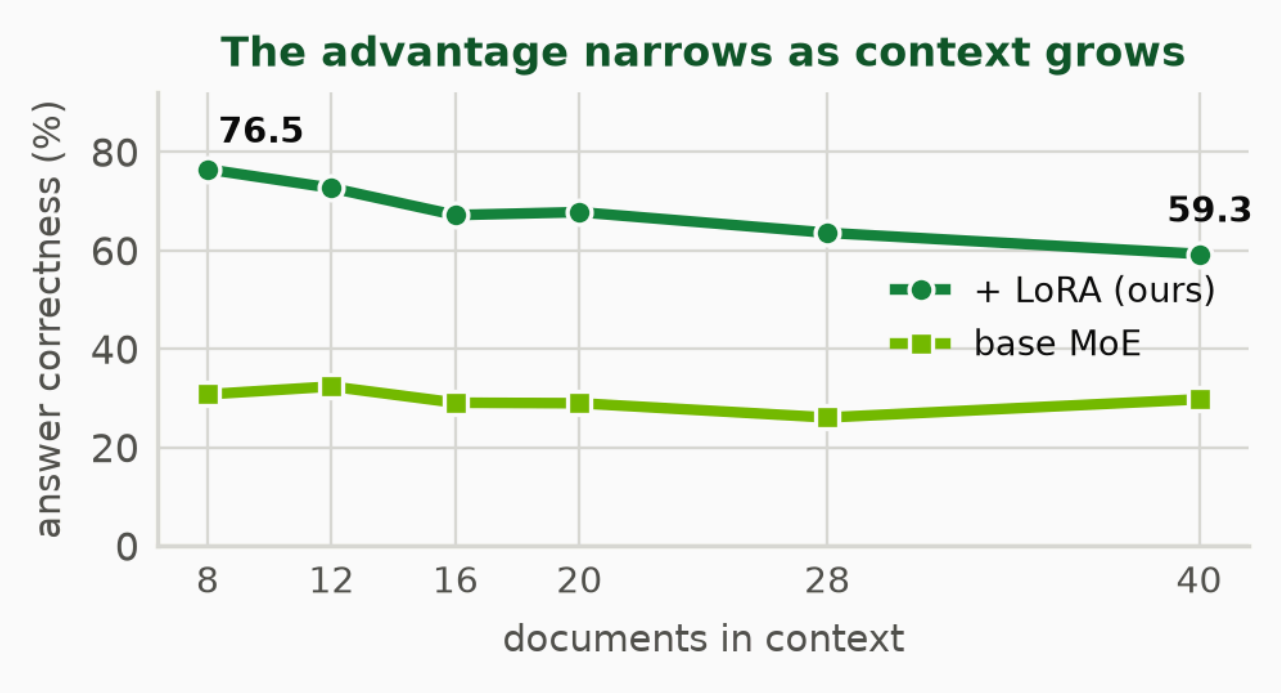

Figure 11: Correctness by context size. The base model's flatness is a floor, not robustness: it is not degrading with context because it was never using it.

**Abstention is the honest caveat.** Refusing unanswerable questions improves from 1.2% to 30.5% over 1,234 unanswerable items ($[0.7, 2.0]$ and $[28.0, 33.1]$), a gain that is still a failing grade. The adapted reader answers most unanswerable questions rather than declining them. Anyone deploying this should treat abstention as unsolved.

This is also the one place the base model wins. False abstention, declining a question that was in fact answerable, rises from **0.4% to 3.4%** ($[0.3, 0.7]$ and $[2.9, 4.1]$). As rates that trade looks ten to one; in items it is nearer three to one, because the answerable partition is three times the larger: roughly 361 additional correct refusals against roughly 112 additional wrong ones. Favourable still, but it is a real regression and we report it rather than let the improving metrics speak alone.

## 10. Four ways our instruments lied

Every one of these produced a plausible, publishable, *wrong* number. None was caught by inspection. Each was caught only when a second instrument disagreed.

1. **Adapting attention alone was worse than not adapting.** On our internal Greek evaluation, attention-only LoRA landed *below* the untuned base model, while adapting all linear layers gained substantially. On hybrid Mamba-Transformer stacks, most of the capacity is not in the attention blocks.

2. **The scorer read a Greek-answering model as noise.** Our MCQ evaluator scored Latin answer letters only. A model that had successfully learned to answer in Greek letters was recorded as random guessing, a fake catastrophic-forgetting result.
3. **A missing prompt prefix made the validation curve run backwards.** The evaluator omitted a prefix the model was trained with, so the validation curve fell steadily while true held-out quality rose.
4. **The distributed loss was summed, not averaged.** Effective learning rate scaled silently with world size, so runs were not comparable across GPU counts.

## 11. What to take away

**Measure your baseline before you buy a bigger model.** On our specialist Greek domains an 8B embedder loses to BM25 and ties a 4B: scale was the wrong axis there, and language exposure was the right one. On general Greek the same family separates cleanly and the 8B does catch BM25, so the lesson is not that scale never helps. It is that you cannot know which axis you are on until you have run the lexical baseline on *your* corpus.

**Always evaluate against the floor, then evaluate again.** A cross-encoder that looked reasonable in isolation contributed nothing over its own first stage on our first evaluation, and a modest, uneven amount on a larger one. Only the no-rerank control could show either, and only the re-run showed how much the first answer depended on the sample.

**Build the evaluation first when the language has none.** Greek had no RAG benchmark, so every claim here would have been unfalsifiable without building one. It also caught the abstention failure that the headline numbers hide.

**Adaptation is cheap, and it is bounded by the domains you adapt on.** A 1B embedder and a 1B reranker, both fine-tuned in hours, beat an off-the-shelf stack by 52% end to end, and a LoRA on a 3B-active MoE more than doubled grounded answer correctness. Off those domains the picture changes: on general Greek Wikipedia our adapted embedder is beaten by BM25 and by an off-the-shelf multilingual model. Adaptation bought language competence, which transfers, and domain fit, which does not.

## Availability

The three models and the benchmark ship as one collection, *Sophea Nemo RAG*: huggingface.co/collections/KIEFERSA/sophea-nemo-rag

`Sophea-Nemo-Embedding` .. embedder
`Sophea-Nemo-Reranker` .. reranker
`Sophea-RAG-Nemo3` .. grounded reader
`HERA` .. reader benchmark, CC-BY-SA-4.0

The reader-supervision teacher `KIEFERSA/Sophea-Titan-1` (Apache-2.0) is published separately. The retrieval training corpus is not part of this release, so the retrieval results here can be reproduced against the released models and benchmark but not retrained from scratch. Base-model licences permit commercial use, and training data was curated to exclude non-commercial sources.

**What we are not claiming.** The reader numbers are single-run point estimates with no intervals, scored by an LLM judge on a benchmark whose human verification is an unsized spot-check and whose judge shares a family with its generator (Section 4), so judged correctness is an upper bound; the retrieval numbers are measured on synthetic queries and overstate production lift (Section 3); that margin is in-domain only and reverses on general-domain Greek; on medical the margin over BM25 is +0.027 with an interval that touches zero, so we do not claim an advantage over lexical search there (Section 6); labels are chunk-as-gold rather than human-annotated; the Greek corpus skews formal and official, so conversational Greek is under-represented; abstention remains poor; and we have not tested hybrid BM25+dense fusion *as a serving first stage* beyond the single fusion configuration reported in Section 6, and we have not tuned $k$, the candidate depth, or the score normalisation at all.

## References

Mikel Artetxe, Sebastian Ruder, and Dani Yogatama. 2020. On the Cross-lingual Transferability of Monolingual Representations. In *Proceedings of ACL 2020*.

Lucas Bandarkar, Davis Liang, Benjamin Muller, Mikel Artetxe, Satya Narayan Shukla, Donald Husa, Naman Goyal, Abhinandan Krishnan, Luke Zettlemoyer, and Madian Khabsa. 2024. The Belebele Benchmark: a Parallel Reading Comprehension Dataset in 122 Language Variants. In *Proceedings of ACL 2024*.

Gordon V. Cormack, Charles L. A. Clarke, and Stefan Buettcher. 2009. Reciprocal Rank Fusion Outperforms Condorcet and Individual Rank Learning Methods. In *Proceedings of SIGIR 2009*, pages 758–759.

Shahul Es, Jithin James, Luis Espinosa-Anke, and Steven Schockaert. 2024. RAGAS: Automated Evaluation of Retrieval Augmented Generation. In *Proceedings of EACL 2024 (System Demonstrations)*.

William Fedus, Barret Zoph, and Noam Shazeer. 2022. Switch Transformers: Scaling to Trillion Parameter Models with Simple and Efficient Sparsity. *Journal of Machine Learning Research*, 23(120):1–39.

Luyu Gao, Yunyi Zhang, Jiawei Han, and Jamie Callan. 2021. Scaling Deep Contrastive Learning Batch Size under Memory Limited Setup. In *Proceedings of the 6th Workshop on Representation Learning for NLP (RepL4NLP)*.

Tianyu Gao, Howard Yen, Jiatong Yu, and Danqi Chen. 2023. Enabling Large Language Models to Generate Text with Citations. In *Proceedings of EMNLP 2023*, pages 6465–6488.

Edward J. Hu, Yelong Shen, Phillip Wallis, Zeyuan Allen-Zhu, Yuanzhi Li, Shean Wang, Lu Wang, and Weizhu Chen. 2022. LoRA: Low-Rank Adaptation of Large Language Models. In *International Conference on Learning Representations (ICLR)*.

Kalervo Järvelin and Jaana Kekäläinen. 2002. Cumulated Gain-based Evaluation of IR Techniques. *ACM Transactions on Information Systems*, 20(4):422–446.

Vladimir Karpukhin, Barlas Oğuz, Sewon Min, Patrick Lewis, Ledell Wu, Sergey Edunov, Danqi Chen, and Wen-tau Yih. 2020. Dense Passage Retrieval for Open-Domain Question Answering. In *Proceedings of EMNLP 2020*.

Aditya Kusupati, Gantavya Bhatt, Aniket Rege, Matthew Wallingford, Aditya Sinha, Vivek Ramanujan, William Howard-Snyder, Kaifeng Chen, Sham Kakade, Prateek Jain, and Ali Farhadi. 2022. Matryoshka Representation Learning. In *Advances in Neural Information Processing Systems (NeurIPS)*.

Nelson F. Liu, Kevin Lin, John Hewitt, Ashwin Paranjape, Michele Bevilacqua, Fabio Petroni, and Percy Liang. 2024. Lost in the Middle: How Language Models Use Long Contexts. *Transactions of the Association for Computational Linguistics*, 12:157–173.

Lefteris Loukas, Nikolaos Smyrnioudis, Chrysa Dikonimaki, Spyros Barbakos, Anastasios Bompotas, Ion Androutsopoulos, Prodromos Malakasiotis, and Ilias Chalkidis. 2025. GR-NLP-TOOLKIT: An Open-Source NLP Toolkit for Modern Greek. In *Proceedings of COLING 2025 (System Demonstrations)*.

Rodrigo Nogueira and Kyunghyun Cho. 2019. Passage Re-ranking with BERT. *arXiv preprint arXiv:1901.04085*.

NVIDIA. 2025b. Nemotron 3 Nano: Open, Efficient Mixture-of-Experts Hybrid Mamba-Transformer Model for Agentic Reasoning. *arXiv preprint arXiv:2512.20848*.

NVIDIA. 2025a. NVIDIA Nemotron 3: Efficient and Open Intelligence. *arXiv preprint arXiv:2512.20856*.

Aaron van den Oord, Yazhe Li, and Oriol Vinyals. 2018. Representation Learning with Contrastive Predictive Coding. *arXiv preprint arXiv:1807.03748*.

Stephen Robertson and Hugo Zaragoza. 2009. The Probabilistic Relevance Framework: BM25 and Beyond. *Foundations and Trends in Information Retrieval*, 3(4):333–389.

Dimitris Roussis, Georgios Paraskevopoulos, Leon Voukoutis, Sokratis Sofianopoulos, Prokopis Prokopidis, Vassilis Papavassiliou, Athanasios Katsamanis, Stelios Piperidis, and Vassilis Katsouros. 2025. Krikri: Advancing Open Large Language Models for Greek. *arXiv preprint arXiv:2505.13772*.

Jon Saad-Falcon, Omar Khattab, Christopher Potts, and Matei Zaharia. 2024. ARES: An Automated Evaluation Framework for Retrieval-Augmented Generation Systems. In *Proceedings of NAACL 2024*.

Noam Shazeer, Azalia Mirhoseini, Krzysztof Maziarz, Andy Davis, Quoc Le, Geoffrey Hinton, and Jeff Dean. 2017. Outrageously Large Neural Networks: The Sparsely-Gated Mixture-of-Experts Layer. In *International Conference on Learning Representations (ICLR)*.

Shivalika Singh, Freddie Vargus, Daniel D'souza, and others. 2024. Aya Dataset: An Open-Access Collection for Multilingual Instruction Tuning. In *Proceedings of ACL 2024*.

Nandan Thakur, Nils Reimers, Andreas Rücklé, Abhishek Srivastava, and Iryna Gurevych. 2021. BEIR: A Heterogeneous Benchmark for Zero-shot Evaluation of Information Retrieval Models. In *NeurIPS Datasets and Benchmarks Track*.

Ahmet Üstün, Viraat Aryabumi, Zheng-Xin Yong, Wei-Yin Ko, Daniel D'souza, Gbemileke Onilude, Neel Bhandari, Shivalika Singh, Hui-Lee Ooi, Amr Kayid, and others. 2024. Aya Model: An Instruction Finetuned Open-Access Multilingual Language Model. In *Proceedings of ACL 2024*.

Leon Voukoutis, Dimitris Roussis, Georgios Paraskevopoulos, Sokratis Sofianopoulos, Prokopis Prokopidis, Vassilis Papavassiliou, Athanasios Katsamanis, Stelios Piperidis, and Vassilis Katsouros. 2024. Meltemi: The First Open Large Language Model for Greek. *arXiv preprint arXiv:2407.20743*.

Liang Wang, Nan Yang, Xiaolong Huang, Linjun Yang, Rangan Majumder, and Furu Wei. 2024. Multilingual E5 Text Embeddings: A Technical Report. *arXiv preprint arXiv:2402.05672*.

Xinyu Zhang, Nandan Thakur, Odunayo Ogundepo, Ehsan Kamalloo, David Alfonso-Hermelo, Xiaoguang Li, Qun Liu, Mehdi Rezagholizadeh, and Jimmy Lin. 2023. MIRACL: A Multilingual Retrieval Dataset Covering 18 Diverse Languages. *Transactions of the Association for Computational Linguistics*, 11:1114–1131.

Yanzhao Zhang, Mingxin Li, Dingkun Long, Xin Zhang, Huan Lin, An Yang, Pengjun Xie, Wen Zhang, and Jingren Zhou. 2025. Qwen3 Embedding: Advancing Text Embedding and Reranking Through Foundation Models. *arXiv preprint arXiv:2506.05176*.

Lianmin Zheng, Wei-Lin Chiang, Ying Sheng, Siyuan Zhuang, Zhanghao Wu, Yonghao Zhuang, Zi Lin, Zhuohan Li, Dacheng Li, Eric P. Xing, Hao Zhang, Joseph E. Gonzalez, and Ion Stoica. 2023. Judging LLM-as-a-Judge with MT-Bench and Chatbot Arena. In *NeurIPS Datasets and Benchmarks Track*.

# A Metric definitions

**Ranking quality.** For a ranked list of $k$ documents with binary relevance $r_i \in \{0,1\}$, discounted cumulative gain and its normalised form are (Järvelin and Kekäläinen, 2002)

$$\mathrm{DCG@}k = \sum_{i=1}^{k} \frac{r_i}{\log_2(i+1)}, \quad \mathrm{nDCG@}k = \mathrm{DCG@}\frac{k}{\mathrm{IDCG}}@k$$

where IDCG@$k$ is the same quantity for the ideal ranking, so nDCG@$k \in [0,1]$. Retrieval figures are macro means of nDCG@10, and Table 2 reports both weightings because they differ: *macro over sets* averages the four evaluation sets equally, *macro over queries* weights each of the 5,830 queries equally. Paired bootstrap intervals are computed per query, so the +0.080 margin over BM25 is the macro-over-queries difference ($0.8575 - 0.7772$); the macro-over-sets difference is +0.078. Elsewhere in the report, absolute levels are quoted macro over sets.

**Out-of-domain runs.** The HERA retrieval track and the medical set of Section 6 were scored with a `max_seq_length` of 4,096 rather than each model's own limit. The HERA corpus contains passages up to 26,000 characters and the models default to 32,768-token windows, which exhausts memory; 4,096 matches our training length and sits above the corpus 99th percentile of roughly 1,870 tokens, so truncation is negligible. Prompt conventions are unchanged.

**Coverage.** With $\mathcal{R}_k$ the top-$k$ retrieved set and $\mathcal{G}$ the gold set for a query,

$$\text{Recall@}k = \frac{|\mathcal{R}_k \cap \mathcal{G}|}{|\mathcal{G}|}, \quad \text{MRR@}k = \frac{1}{|Q|}\sum_{q\in Q}\frac{1}{\text{rank}_q}$$

where $\text{rank}_q$ is the position of the first relevant document and the term is 0 if none appears within $k$. Recall@$k$ is the quantity that bounds the reader (Section 8).

**Lexical baseline.** BM25 scores a query–document pair as (Robertson and Zaragoza, 2009)

$$\sum_{t\in q}\text{IDF}(t)\cdot\frac{f_{t,d}(k_1+1)}{f_{t,d}+k_1\left(1-b+b\ |d|\ /\overline{|d|}\right)}$$

with $f_{t,d}$ the frequency of term $t$ in document $d$, $|d|$ its length, $\overline{|d|}$ the mean document length, and $k_1 = 1.2$, $b = 0.75$. Tokenisation is Greek-aware: NFD normalisation, combining marks stripped, final sigma folded.

**Grounded reading.** Let $\mathcal{C}$ be the citation set a reader emits for an item and $\mathcal{G}$ the gold set. Following the ALCE decomposition (Gao et al., 2023),

$$P = \frac{|\mathcal{C}\cap\mathcal{G}|}{|\mathcal{C}|}, \quad R = \frac{|\mathcal{C}\cap\mathcal{G}|}{|\mathcal{G}|}, \quad F_1 = \frac{2PR}{P+R}$$

and exact citation-set match is $\mathbf{1}[\mathcal{C}=\mathcal{G}]$, a strictly harder criterion than $F_1$ because a single spurious index fails the item outright.

Over the unanswerable and answerable partitions of the benchmark,

$$\text{abstention recall} = \frac{\#\{\text{unanswerable, refused}\}}{\#\{\text{unanswerable}\}},$$

$$\text{false abstention} = \frac{\#\{\text{answerable, refused}\}}{\#\{\text{answerable}\}}.$$

Answer correctness and faithfulness are not closed-form: both are judged by a served LLM, correctness against the reference answer on grounded-and-answered items, faithfulness by asking whether every claim in the response is supported by the retrieved context. Faithfulness is therefore scored against the context rather than against any reference text, which is why Section 4 treats it as the metric least exposed to the generator confound.

# B What each number was measured on

The reported comparisons do not all share a query set or a first stage, so they are not interchangeable. This table is the reconciliation.

**Embedder** (Table 2, Figures 1–6). 5,830 held-out queries, corpus = all positives. No maximum-length override is applied at scoring time, so each model runs at its own configured limit. Queries carry the Qwen3-Embedding instruction convention for that family; Nemotron-3-Embed is prompted on *both* sides (`::query: / ::passage:`) because its mean-pool includes the prompt.

**Reranker, first evaluation** (Table 5, Figure 7). 150 queries per set, 750 pooled, seed 42; rerank maximum length 2,048. The fixed first stage is *our fine-tuned Qwen3-0.6B embedder*, not the adapted Nemotron 1B, so the floor here is not a row of Table 2 and the +0.044 is not directly composable with the embedder numbers. The off-the-shelf arm is `llama-nemotron-rerank-1b-v2`.

**Reranker, second evaluation** (Table 6). 2,580 queries: 593 validation, 263 energy, 879 legal, 195 finance and the 650-query medical set of Section 6. Same fixed first stage, same top-50 candidate depth, same rerank maximum length, same seed. Each set is scored against its own positives-plus-negatives corpus (693 to 4,780 passages), so absolute levels are not comparable to the first evaluation or to Table 2; the quantity carried across runs is the paired difference against the floor, which is computed within a run.

**Chained** (Figure 9). The same 750 pooled queries, both stages swapped together, off-the-shelf Nemotron embedder and reranker against both of ours. Means are macro over the five sets.

**Reader** (Section 9, Figure 10 and Figure 11). All 4,946 HERA items, 3,712 answerable and 1,234 unanswerable, on gold-provided context.

# C The HERA metric set

HERA is released with three scoring tracks, and they are deliberately separable: a system can fail at retrieval, at reading, or only when the two are chained, and the tracks are designed to tell those cases apart. This report exercises the reader track and the robustness slice.

## C.1 Retrieval track

Scored against each item's `gold_ids` over `corpus.jsonl`, with no reader involved.

**Recall@*k* and Hit@*k*** measure whether the gold passage is in the top $k$ at all. This is the track's most consequential number, because it is the bound of Section 8: a passage that never enters the context cannot be read, so Recall@$k$ caps every reader metric downstream regardless of reader quality.

**nDCG@*k* and MRR** measure how well the gold passage is placed once found. They differ in what they reward: nDCG credits every relevant passage with a rank-discounted weight, so it responds to multi-gold items, while MRR looks only at the first relevant hit and is therefore the better proxy for a reader that reads the top document and stops.

## C.2 Reader track

Scored on gold-provided context, so that reading is measured without retrieval error mixed in.

**Answer correctness** is semantic agreement with the gold answer, graded by an LLM judge rather than by exact match. Exact match is unusable here: answers are free-form Greek, and a correct paraphrase with different morphology would be scored wrong.

**Faithfulness** asks whether every claim in the response is supported by the supplied context. It is deliberately independent of correctness, because the two failures come apart. A model can be correct but ungrounded, reciting a fact it knew rather than one it read, and it can be faithful but wrong, misreading a passage it genuinely used. Only faithfulness detects the first, which is the failure a retrieval system exists to prevent.

**Citation F1** compares the context positions the reader cites against the gold `citations` positions. Precision penalises citing everything, which would otherwise be a trivial way to guarantee recall; recall penalises answering without attribution.

**Abstention** is scored on the two partitions separately: recall on the unanswerable items, i.e. how often the model correctly refuses, and the false-abstention rate on the answerable ones. Both are needed, since either alone is gameable by a model that always refuses or never does.

### C.3 End-to-end and robustness

**End-to-end quality** re-runs the reader metrics on *retrieved* rather than gold-provided context. This is the only number that reflects deployment, and it is where retrieval and reading errors compound.

**Positional robustness** reports accuracy as a function of `n_docs` and of where the gold passage sits in the context, which is what exposes the lost-in-the-middle effect (Liu et al., 2024) and what Figure 11 and the position breakdown in Section 9 are drawn from.

## D Generation and judging configuration

**Query synthesis.** Temperature 0.7, top-$p$ 0.9, `max_new_tokens` 256 (512 for the few-shot domain prompts), seed 42, two or three queries per chunk, request concurrency 32 against an OpenAI-compatible vLLM endpoint.

**Judging.** Temperature 0, one judge call per axis, output constrained to a single JSON object so the verdict is parsed rather than interpreted. The judge endpoint is probed live at startup and the first reachable one is used, because served aliases drift.

**Thinking must be disabled, and this is not optional.** Every generator and judge here is a reasoning model. Called with its default chat template, the model spends the entire token budget on a reasoning trace and returns *zero* parseable JSON: `finish_reason` comes back as `length` with empty content. The fix is to pass

```
chat_template_kwargs={"enable_thinking": false}
```

at the request layer. A `"detailed thinking off"` system message does *not* work, and neither does a bare `thinking: false` key. Disabling thinking also made synthesis roughly ten times faster with no measurable quality loss.

## E Prompts

The prompts below are English translations; the originals are Greek, and the exact strings ship with the evaluation scripts.

**Reader system prompt.** Identical for the base and adapted models, which is what makes the citation and abstention rows in Figure 10 a measure of compliance rather than of instruction:

> You are an assistant that answers EXCLUSIVELY on the basis of the supplied documents. Use only the information in the documents; do NOT invent. Cite in brackets the number of the document or documents you used, e.g. [2]. If the answer is NOT present in the documents, say so plainly. Answer in Greek, briefly and to the point.

The user turn is the numbered context, one block per passage as `[n] text`, followed by the question.

**Judge prompts.** One per axis, each returning a single JSON field.

> *Abstention.* You will see an assistant RESPONSE. Say whether the assistant REFUSES to answer because the information is not in the documents (`abstain`), or whether it ATTEMPTS a substantive answer (`answer`). JSON only: `{"verdict":"abstain|answer"}`

> *Correctness.* You are an evaluator. You will see a QUESTION, a SOURCE PASSAGE (the correct information) and an assistant RESPONSE. Judge whether the response is substantively CORRECT according to the source passage. JSON only: `{"correct": true|false}`

> *Faithfulness.* You are a faithfulness evaluator. You will see DOCUMENTS and a RESPONSE. Judge whether EVERY claim in the response is supported by the documents (no invention or hallucination). JSON only: `{"faithful": true|false}`